\documentclass[a4paper,11pt]{article}
\usepackage{jheppub} 
\usepackage{lineno}
\usepackage{subfigure}

\title{Gluon-gluon fusion contribution to the productions of three gauge bosons at the LHC}

\author[a,b]{Jianpeng Dai,}
\author[c,b]{Zhenghong Hu,}
\author[c,b]{Tao Liu,}
\author[a,b]{Jin Min Yang}
 
\affiliation[a]{CAS Key Laboratory of Theoretical Physics, Institute of Theoretical Physics,
Chinese Academy of Sciences, Beijing 100190, P. R. China }
\affiliation[b]{School of Physical Sciences, University of Chinese Academy of Sciences, Beijing 100049, P. R. China}
\affiliation[c]{Institute of High Energy Physics, Chinese Academy of Sciences, Beijing 100049, P. R. China}

\emailAdd{daijianpeng@mail.itp.ac.cn}
\emailAdd{huzh@ihep.ac.cn} 
\emailAdd{liutao86@ihep.ac.cn}
\emailAdd{jmyang@itp.ac.cn}

\abstract{Productions of multiple gauge bosons at the LHC are sensitive to triple or quartic gauge couplings and thus provide a sensitive test for the electroweak sector of the Standard Model and allow for a probe of new physics.   
In this work we calculate the gluon-gluon initiate state contribution to the productions of three gauge bosons ($Z\gamma\gamma$, $ZZ\gamma$ and $W^+W^-\gamma$) at the LHC, which is formally part of NNLO effects compared  to the LO quark-antiquark channels corrections. For each process we present the ratio between the gluon-gluon channels contribution and the quark-antiquark channels contribution. 
We found that such a ratio for $Z\gamma\gamma$ ($ZZ\gamma$) is of the order of $10^{-3}$ ($10^{-4}$), much smaller than the corresponding ratio for the diboson production due to the decrease of gluon PDF when more particles appear in the final states.  
These small ratios imply that gluon-gluon fusion contribution is phenomenological negligible for the productions of $Z\gamma\gamma$ and $ZZ\gamma$. However, for $W^+W^-\gamma$ production, the ratio is about 5\%, which is of the same order of magnitude as the ratio for $W^+W^-$ production due to the big cancellation between the amplitudes of quark-antiquark channels. While such an effect can be neglected currently at the LHC,  it may be accessible at the HL-LHC. }

\begin{document}
\maketitle
\flushbottom

\section{Introduction}
\label{sec:intro}

Productions of multiple gauge bosons at the LHC are sensitive to triple or quartic gauge couplings at tree level of scattering amplitudes, and thus provide a sensitive test for the electroweak (EW) sector of the Standard Model (SM) besides vector boson scattering (VBS)  processes~\cite{Covarelli:2021gyz}. Any deviation from the SM prediction would be an indication of new physics beyond the SM (BSM). Also, they could be important backgrounds for many SM and BSM processes, e.g., $W/Z$ boson plus two photons for Higgs production in association with $W/Z$ boson where the Higgs decays to two photons. In contrast to diboson productions, triboson processes are generally quite rare if the leptonic decay channels are considered (the hadronic final states would have huge QCD backgrounds at hadron colliders). Recently, ATLAS and CMS observed some productions of three gauge bosons for the first time from proton-proton collisions with an unprecedented integrated luminosity, such as the productions of three massive gauge bosons ~\cite{ATLAS:2019dny,CMS:2020hjs,ATLAS:2022xnu,CMS:2019mpq}, one massive plus two massless photons~\cite{CMS:2021jji,ATLAS:2022wmu,ATLAS:2023avk}, 
and two massive plus one massless photon~\cite{ATLAS:2023zkw,CMS:2023jpy}. 
On the other hand, the SM Lagrangian is expanded to include high dimensional operators to parameterize BSM effects in the SM effective field theory (SMEFT)~\cite{Grzadkowski:2010es,Murphy:2020rsh,Li:2020gnx}, which provides a convenient way to understand correlations between various experimental results and has been widely used in both experimental and theoretical studies. 
Some analyses for the diboson, triboson and VBS processes have been performed in the framework of SMEFT ~\cite{Green:2016trm, Zhang:2016zsp,Baglio:2017bfe,Ellis:2018gqa,Baglio:2019uty,Bellan:2021dcy,Senol:2021wza,Degrande:2023iob,Bellan:2023efn}.

Before discussing triboson productions at the LHC, we first take a look at diboson productions. It was found that the gluon-gluon initial state channels could contribute ${\cal O}(10\%)$ to the leading order cross section which comes from the quark-antiquark channels~\cite{Campbell:2011bn}, if the total charge of the produced diboson vanishes, i.e., $\{\gamma\gamma\,~Z\gamma,~ZZ,~W^+W^-\}$. All the external particles in the gluon-gluon channels are connected to a closed fermion loop and they are formally next-to-next-to-leading order (NNLO) corrections, while the large gluon flux in the parton distribution function (PDF) would compensate the loop factor $({\alpha_s}/{\pi})^2$ suppression. Then for triboson productions it is also expected that there may be similar non-negligible contributions from gluon-gluon fusion, which is one motivation of this work. We will evaluate the contribution of gluon-gluon fusion to the neutral-charge production processes, $gg\rightarrow \{Z\gamma\gamma\,~ZZ\gamma,~W^+W^-\gamma\}$ at the parton level. 
The NLO QCD corrections to such processes from quark-antiquark channels with leptonic decays can be found in~\cite{Hankele:2007sb,Binoth:2008kt,Bozzi:2009ig,Bozzi:2011en}. Electroweak (EW) contributions at NLO have also been evaluated in ~\cite{Nhung:2013jta,Shen:2015cwj,Wang:2016fvj,Greiner:2017mft,Zhu:2020ous}. Here we refrain from providing a review on high order corrections to triboson productions but refer to~\cite{Huss:2022ful} for more details.
Since there are no technical problems for evaluating one-loop five-point Feynman integrals, in our analysis we will try to understand the numerical results through their relations with diboson production at the LHC. 

Furthermore, considering the amplitudes of $gg\rightarrow \gamma\gamma\gamma$, we know from Furry theorem that there is at least one axial-vector coupling for each Feynman diagram to have non-vanishing effects. It means that the triphoton amplitudes will have an overall anti-symmetric tensor $\epsilon_{\mu\nu\rho\sigma}$, which first appears at the two-loop level. As for the massive triboson productions, the needed axial vector couplings can appear at the leading one-loop level. Since such axial-vector couplings are not necessary to appear for diboson productions in gluon fusion channels, the calculation of triboson productions may be quite different from diboson productions. So, an explicit calculation of triboson productions in gluon fusion channels is necessary for a phenomenological analysis.

This work is organized as follows. 
In the next section some details of the calculation will be described and the results will be shown in three subsections. Finally, the conclusion is made in Section 3. 

\section{Calculations and results}

In our calculation we use {\tt MadGraph5\_aMC@NLO} with version 3.4.2~\cite{Alwall:2014hca} for Monte Carlo simulations. We also use {\tt FeynArts} and {\tt FormCalc}~\cite{Hahn:1998yk,Hahn:2000kx,Hahn:2006qw} to cross-check and to get the detailed information of the physical amplitudes. 
Due to the numerical instability problem caused by the inverse Gram determinants in the conventional Passarino-Veltman reduction~\cite{tHooft:1978jhc}, we adopt the reduction scheme proposed in~\cite{Denner:2002ii,Denner:2005nn} for one-loop five-point tensor integrals, which has been implemented in the public code {\tt Collier}~\cite{Denner:2016kdg}. For parton distribution functions, we use LHAPDF6~\cite{Buckley:2014ana} with NNPDF3.0 set~\cite{NNPDF:2014otw} at LO (with $\alpha_s(m_Z)=0.1247$) and NNLO (with $\alpha_s(m_Z)=0.1190$) fit for quark channels and gluon channels respectively. 

The central values of factorization and renormalization scales are set to be the same as the dynamical partonic center-of-mass energy, $\mu_R^0=\mu_F^0 = \sqrt{\hat s}$. 
And we perform a 7-point scale variation to estimate the scale uncertainty. The relevant parameters used in the evaluation are   
\begin{alignat}{2}
    m_b &= 4.7~\mathrm{GeV}, &~~~\quad m_Z &= 91.188~\mathrm{GeV}, \nonumber\\ 
    m_t &= 173~\mathrm{GeV}, &~~~\quad m_W &= 80.419~\mathrm{GeV}, \\ 
    m_h &= 125~\mathrm{GeV}, &~~~\quad G_F &= 1.16639\times 10^{-5}~\mathrm{GeV^{-2}}, \nonumber \\
    \alpha &= \frac{1}{132.507}. \nonumber
\end{alignat}
Other quarks not listed above are thought to be massless. The collision energy $\sqrt{s}$ is set to be $13~\rm TeV$.  We use the following basic cuts for photons: 
\begin{equation}
    p^\gamma_T > p^\gamma_{T,min},~\left| \eta^\gamma  \right| < 2.37,~\Delta R_{\gamma \gamma} > 0.4.
    \label{Eq1}
\end{equation}
But we do not apply any cuts on massive vector bosons ($Z$ and $W^\pm$).
Here $p^\gamma_{T,min}$ is chosen as a free parameter to see its impact on the total cross sections.

\subsection{\texorpdfstring{$Z \gamma \gamma$}{Zyy} and \texorpdfstring{$Z\gamma$}{Zy} productions}

We start with $pp\rightarrow Z \gamma \gamma$, which was measured by the CMS and ATLAS Collaborations recently~\cite{ATLAS:2022wmu,CMS:2021jji}. These two experimental groups both used {\tt Madgraph5\_aMC@NLO} in their analysis. 
Here we choose $Z\gamma$ production as the reference process for comparison, since it is naively expected that the phase space would not change much by an additional photon and hence the gluon initiated channels may provide contribution of same order of magnitudes for both diboson and triboson productions, i.e. $\sigma^{gg}(Z\gamma\gamma)/\sigma^{q\bar q}(Z\gamma\gamma) \simeq  \sigma^{gg}(Z\gamma)/\sigma^{q\bar q}(Z\gamma)$. High order corrections to $Z\gamma$ and $Z\gamma\gamma$ productions at the LHC have also been calculated in various directions, e.g., the NLO corrections to $pp\rightarrow Z \gamma \gamma$ with the leptonic decays of $Z$-boson have been studied in~\cite{Bozzi:2011en}. 
Since we are only interested in the ratio $\sigma^{gg}/\sigma^{q\bar q}$, for simplicity only the tree-level contribution of the quark-antiquark channels and the one-loop contribution of the gluon-gluon channels are considered in the following analysis.      

\begin{figure}[htbp]%
    \centering
    \subfigure[]{
        \includegraphics[width=10cm]{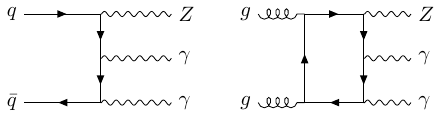}
        }\hfill \\
    \subfigure[]{
        \includegraphics[width=10cm]{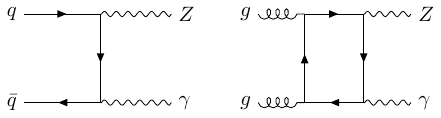}
        }
    \caption{Typical Feynman diagrams that contribute to (a) $q\Bar{q} \rightarrow Z\gamma\gamma$ and $gg\rightarrow Z\gamma\gamma$,  (b) $q\Bar{q} \rightarrow Z\gamma$ and $gg\rightarrow Z\gamma$.}
    \label{fig:Fig1}
\end{figure}
\begin{figure}
    \centering
    \includegraphics[width=10.5cm]{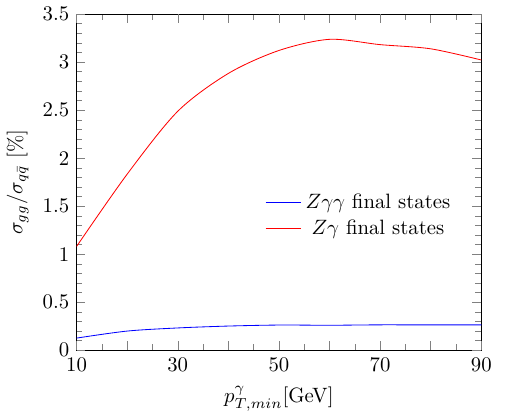}
    \caption{The ratio $\sigma^{gg}/\sigma^{q\Bar{q}}$ for $Z\gamma\gamma$ and $Z\gamma$ productions at the LHC with $\sqrt{s}=13$ TeV, as a function of $p^\gamma_{T,min}$.}
    \label{fig:Fig2}
\end{figure}
\begin{figure}
    \centering
    \includegraphics[width=15cm]{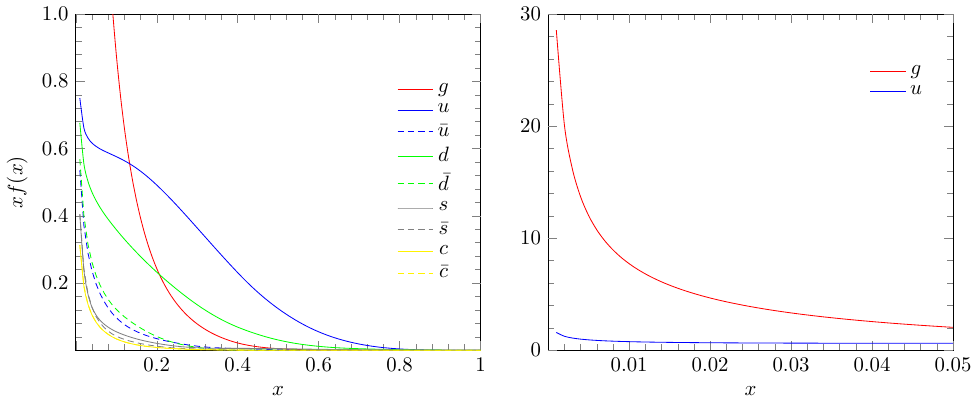}
    \caption{Left: Parton distribution functions $xf(x)$ for quarks, antiquarks, and gluons in the proton. Right: Parton distribution functions $xf(x)$ for gluons and $u$ quark at small momentum fractions. These values are obtained from NNPDF3.0~\cite{NNPDF:2014otw} at $Q=m_Z$.}
    \label{fig:Fig3}
\end{figure}

The typical Feynman diagrams contributing to $pp\rightarrow \{ Z\gamma,~Z\gamma\gamma \}$ are shown in Fig.~\ref{fig:Fig1}. Total cross sections at different $p^\gamma_{T,min}$ are given in Table~\ref{tab:tab1}. In Fig.~\ref{fig:Fig2} we show the ratio of $\sigma^{gg}$ to $\sigma^{q\Bar{q}}$ as a function of $p^\gamma_{T,min}$. 
It is easy to find that quark-antiquark channels dominates in the low $p^\gamma_{T}$ region, then the ratios $\sigma^{gg}/\sigma^{q{\bar q}}$ reach maximum values at moderate values of $p^\gamma_{T}$. Although not shown explicitly in Fig.~\ref{fig:Fig2}, the ratio for $Z\gamma\gamma$ production also decreases when $p^\gamma_{T}$ gets large, which could be easily found in Table~\ref{tab:tab1}. This behavior is totally determined by the quark and gluon PDFs at the hadron collider, 
which could be easily checked numerically.  The PDFs of the particles which are phenomenological important for the evaluation are shown in Fig.~\ref{fig:Fig3}. 

\begin{table}
    \renewcommand{\arraystretch}{1.5}
    \centering
    \scalebox{0.9}{
    \begin{tabular}{|c|c|c|c|c|}
    \hline
     $p^\gamma_{T,min}~\rm[GeV]$  & $\sigma^{q\Bar{q}}(Z\gamma)~\rm [pb]$ & $\sigma^{gg}(Z\gamma)~\rm [pb]$ &  $\sigma^{q\Bar{q}}(Z\gamma\gamma)~\rm [fb]$ & $\sigma^{gg}(Z\gamma\gamma)~$[$\times10^{-2}~$fb]\\
      \hline
     10  & $75.9(2)^{+8.7\%}_{-9.7\%}$ & $0.818(3)^{+22.9\%}_{-17.0\%}$ & $160.6(6)^{+6.2\%}_{-7.1\%}$ & $20.35(3)^{+22.3\%}_{-16.6\%}$ \\
     20  & $31.15(8)^{+7.4\%}_{-8.4\%}$ & $0.577(1)^{+22.6\%}_{-16.8\%}$ & $41.8(2)^{+4.1\%}_{-4.9\%}$ & $8.26(1)^{+21.4\%}_{-16.5\%}$ \\
     30  & $16.09(4)^{+6.3\%}_{-7.2\%}$ & $0.3986(9)^{+22.2\%}_{-16.6\%}$ & $17.26(5)^{+2.8\%}_{-3.5\%}$ & $4.01(2)^{+22.1\%}_{-17.3\%}$ \\
     40  & $9.35(3)^{+5.4\%}_{-6.3\%}$ & $0.2684(9)^{+21.9\%}_{-16.4\%}$ & $8.89(3)^{+1.8\%}_{-2.3\%}$ & $2.223(3)^{+23.2\%}_{-17.9\%}$ \\
     50  & $5.82(1)^{+4.6\%}_{-5.4\%}$ & $0.1815(9)^{+21.6\%}_{-16.5\%}$ & $5.16(2)^{+1.0\%}_{-1.4\%}$ & $1.346(4)^{+23.9\%}_{-18.3\%}$ \\   
     60  & $3.84(1)^{+3.9\%}_{-4.6\%}$ & $0.1234(3)^{+21.3\%}_{-16.9\%}$ & $3.30(1)^{+0.3\%}_{-0.7\%}$ & $0.869(2)^{+24.9\%}_{-18.9\%}$ \\
     70  & $2.640(7)^{+3.2\%}_{-4.0\%}$ & $0.0843(2)^{+22.0\%}_{-17.2\%}$ & $2.258(9)^{+0.0\%}_{-0.2\%}$ & $0.5962(9)^{+25.6\%}_{-19.3\%}$ \\
     80  & $1.880(3)^{+2.7\%}_{-3.3\%}$ & $0.0591(1)^{+22.5\%}_{-17.5\%}$ & $1.628(7)^{+0.3\%}_{-0.6\%}$ & $0.4280(3)^{+26.2\%}_{-19.6\%}$ \\
     90  & $1.374(3)^{+2.1\%}_{-2.8\%}$ & $0.04142(9)^{+23.0\%}_{-17.8\%}$ & $1.206(5)^{+0.8\%}_{-1.0\%}$ & $0.3179(3)^{+26.6\%}_{-19.9\%}$ \\
     200 & $0.1207(3)^{+1.2\%}_{-1.4\%}$ & $0.002187(4)^{+27.1\%}_{-20.1\%}$ & $0.1421(7)^{+3.7\%}_{-3.5\%}$ & $0.03513(5)^{+29.5\%}_{-21.4\%}$ \\
     300 & $0.02830(7)^{+3.2\%}_{-3.1\%}$ & $0.000350(2)^{+29.0\%}_{-21.1\%}$ & $0.0410(2)^{+5.3\%}_{-4.9\%}$ & $0.00770(2)^{+30.8\%}_{-22.2\%}$ \\
      \hline  
    \end{tabular}   }
    \caption{Total cross sections for $pp \rightarrow \{ Z\gamma,Z\gamma\gamma \}$ at different $p^\gamma_{T,min}$, using the cuts of Eq.~(\ref{Eq1}) for photons. The superscript of $\sigma$ represents different channels. The Monte Carlo integration errors are shown in parentheses and the scale uncertainties are shown in superscript and subscript.}
    \label{tab:tab1}
\end{table}

Another direct observation from the results is that the ratio for $Z\gamma$ production is about 10 times larger than that for $Z\gamma\gamma$. Then we need to   understand why $\sigma^{gg}(Z\gamma\gamma)$ is so small. As mentioned in the introduction, the amplitudes of $gg \rightarrow Z\gamma\gamma$ and $gg \rightarrow Z\gamma$ are totally different from each other. With the help of $C$-parity, one knows that the axial vector interaction between $Z$-boson and quarks only contributes to the former amplitude, while the vector part fully devotes to the latter. Through calculating each one-loop Feynman diagram separately, we find that at the amplitude level the contribution from axial vector part is even larger than the vector part in the process of $Z\gamma\gamma$ production. From the Feynman rules of $Z$-boson couplings to up-type quarks $\frac{g}{4 \cos{\theta_W}} \gamma^\mu \left(1 - \frac{8}{3} \sin^2{\theta_W} -  \gamma_5\right)$ and to down-type quarks $\frac{g}{4 \cos{\theta_W}} \gamma^\mu \left(-1 + \frac{4}{3} \sin^2{\theta_W} + \gamma_5\right)$, one can directly see that the axial related coupling is larger than the other one in the bracket.     

As is well known, top quark only provides sizable contribution to the diboson and triboson amplitudes under the condition of high invariant masses of the final states. Thus, we can only consider the effects of light quarks in the analysis. From the above arguments, it is rather easy to see that the amplitudes of $gg\rightarrow Z\gamma\gamma$ are proportional to $Q_q^2A_q$, where $Q_q$ denotes the electric charge of the quark and $A_f$ represents the axial vector coupling between quarks and $Z$-boson. The fact $A_u= -A_d$ leads to a cancellation between up-type and down-type quark loops. On the other hand, the vector interaction with $Z$-boson parameterized by $V_q$ provides non-vanishing amplitudes $gg\rightarrow Z\gamma$ which are proportional to $Q_qV_q$. And $Q_qV_q$ has the same sign for all the quarks. In order to exclude possible internal cancellations that happen between Feynman diagrams with different ordering of the external legs, we define a new parameter $R(q)=\sigma^*(q)/\sigma(q)$, where $\sigma(q)$ is the ordinary cross section while the amplitudes in $\sigma^*(q)$ are replaced by their absolute values for each Feynman diagram and all other parts in $\sigma^*$ are exactly the same as in $\sigma$. 
Here $q$ denotes the corresponding quark loops in the calculation. Of course, only axial vector interactions with $Z$-boson are considered for $Z\gamma\gamma$ and vector interactions for $gg\rightarrow Z\gamma$. 
The values of $R(q)$ at $p^\gamma_{T,min}=50~\rm GeV$ are shown in Table~\ref{tab:tab2}. These numerical results confirm the above analysis since $(|Q_u|^2 + |Q_d|^2)^2/(Q_u^2 - Q_d^2)^2$ is just equal to the ratio $R(u,d)/R(u)$. We could also find that the degree of cancellation is similar for these two processes, when only one type of quarks are taken into account. Obviously, the cancellation between different quarks can not explain why the ratio $\sigma^{gg}/\sigma^{q\Bar{q}}$ is so suppressed for $Z\gamma\gamma$ production. 

\begin{table}
    \renewcommand{\arraystretch}{1.5}
    \centering
    \scalebox{1.0}{
    \begin{tabular}{|c|c|c|}
    \hline
      final states & $R(u)$ & $R(u , d)$ \\
      \hline
      $Z\gamma$ & $15.87 $& $ 15.87 $ \\
      $Z\gamma\gamma$& $ 14.28 $ & $ 39.68 $\\
      \hline
    \end{tabular} }
    \caption{The value of $R(q)$ for $gg\rightarrow Z\gamma\gamma$ and $gg\rightarrow Z\gamma$ at $p^\gamma_{T,min}=50~\rm GeV$.}
    \label{tab:tab2}
\end{table}

\begin{figure}
    \centering
    \includegraphics[width=10.5cm]{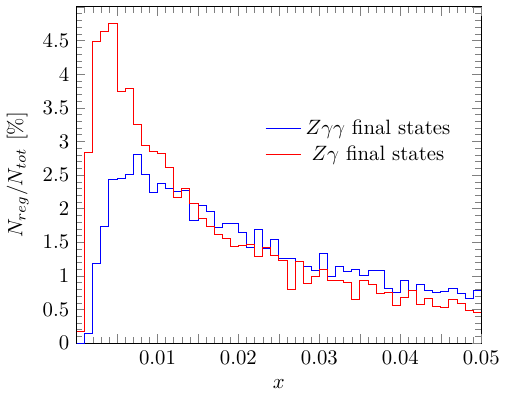}
    \caption{The ratio $N_{reg}/N_{tot}$ for $gg\rightarrow Z \gamma$ and $gg\rightarrow Z\gamma\gamma$ in different momentum fraction regions, using the cuts in Eq.~(\ref{Eq1}) with $p^\gamma_{T,min}=10~\rm GeV$. 
    }
    \label{fig:Fig4}
\end{figure}

Now we show the effects of PDFs. We generate 10000 events for $Z\gamma\gamma$ and $Z\gamma$ by {\tt MadGraph5}, and then use {\tt MadAnalysis5}~\cite{Conte:2012fm} to get the event numbers $N_{reg}$ in different bins of the momentum fraction. The ratio $N_{reg}/N_{tot}$ ($N_{tot}$ is the number of total events) is shown in Fig.~\ref{fig:Fig4} with $p^\gamma_{T,min}=10~\rm GeV$ as an example. Here the distribution for $Z\gamma$ is concentrated in the low fraction region with a peak around $x=0.004$. When an extra photon is added in the final states, the shape of the ratio distribution becomes more flat and the peak moves to $x=0.007$. 
From Fig.~\ref{fig:Fig3} we see that $f(0.004)\simeq \frac{5}{2}f(0.007)$. And in contrast to gluon, there is little change to the quark PDF in the region of small $x$. So we can conclude that the difference of the ratios $\sigma^{gg}/\sigma^{q\Bar{q}}$  shown in Fig.~\ref{fig:Fig2} is mainly due to the suppressed gluon PDF for $Z\gamma\gamma$ production.    

\subsection{\texorpdfstring{$ZZ\gamma$}{ZZy} and \texorpdfstring{$ZZ$}{ZZ} productions}
Compared to $Z\gamma\gamma$ production, the $ZZ\gamma$ production is harder to measure at the LHC due to its lower production rate and the extra suppression factor of $Z$-boson decay. Although there are no published experimental results till now, it is still considered in this work for completeness. Following the same logic as in the preceding subsection, we choose $ZZ$ as its references process with the following LO cross section at the 13 TeV LHC:  
\begin{equation}
    \label{eq2}
    \begin{aligned}
       \sigma^{q\Bar{q}}(ZZ)&=10.98(2)^{+3.0\%}_{-3.8\%}~\rm pb, \\
     \sigma^{gg}(ZZ)&=0.9344(6)^{+21.1\%}_{-17.1\%}~\rm pb. 
    \end{aligned}
\end{equation}
Naively, one would expect that the ratio $\sigma^{gg}(ZZ)/\sigma^{q\Bar{q}}(ZZ)$, which is approximately equal to $9\%$, should be much smaller than the ratio for  the $Z\gamma$ production. Compared to the $gg\rightarrow Z\gamma$ production, an  extra massive $Z$-boson requires a larger $x$ for gluon PDF 
and thus would reduce the total cross section. Seemingly there is a contradiction between the numerical results and our arguments.

\begin{figure}
    \centering
    \includegraphics[width=10cm]{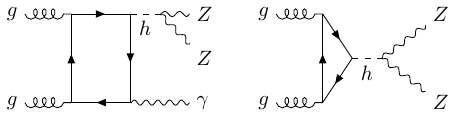}
    \caption{Additional Feynman diagrams contributing to $pp\rightarrow \{ZZ\gamma,ZZ\}$ besides Fig.~\ref{fig:Fig1} where one photon is replaced by a $Z$-boson.}
    \label{fig:Fig6}
\end{figure}

To understand the above puzzle, a close look at the amplitudes is necessary. 
First, after replacing one photon with $Z$-boson in Fig.~\ref{fig:Fig1}, one obtains the corresponding $ZZ\gamma$ production and $ZZ$ production Feynman diagrams. 
Other diagrams which contain Higgs propagators are shown in Fig.~\ref{fig:Fig6}. Obviously, the amplitude for triboson production in this figure vanishes due to $C$-parity. Then the amplitudes of $gg\rightarrow ZZ\gamma$ should have similar structures as $gg\rightarrow Z\gamma\gamma$. The only difference comes from the coupling constants, which are proportional to $Q_qV_qA_q$ and cannot bring significant change to the total cross section. 
About the right diagram of Fig~\ref{fig:Fig6}, seemingly its amplitude should be suppressed by the heavy top quark mass. 
But in real calculations, at least one quark mass has to be picked out in the numerator from the fermion propagators and so no quark mass is left at the leading approximation of the amplitudes. The same property has also been observed in the processes of single and double Higgs productions at the LHC. 
Although this extra amplitude will not be suppressed by the heavy quark mass, it is found that this contribution to the total cross section is small and cannot balance the effect of gluon PDF from numerical calculations. The real reason for relative large $\sigma^{gg}(ZZ)/\sigma^{q\Bar{q}}(ZZ)$ is that the amplitudes of $gg\rightarrow Z\gamma$ are proportional to $Q_qV_q$ and the corresponding $ZZ$ amplitudes without Higgs propagators are proportional to $V_q^2+ A_q^2$. The factor $V_q^2+ A_q^2$ in $ZZ$ production leads to about a factor of 10 enhancement to the cross section compared with $Z\gamma$, which just compensates the suppression by gluon PDF.  
As for the $ZZ\gamma$ production, since there are no such an enhancement at the amplitude level, the ratio $\sigma^{gg}(ZZ\gamma)/\sigma^{q\Bar{q}}(ZZ\gamma)$ should remain small as expected. 

\begin{figure}
    \centering
    \includegraphics[width=10.5cm]{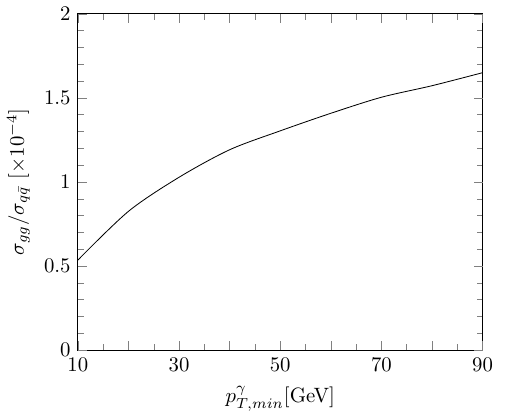}
    \caption{The ratio $\sigma^{gg}(ZZ\gamma)/\sigma^{q\Bar{q}}(ZZ\gamma)$ at  the 13 TeV LHC as a function of $p^\gamma_{T,min}$.}
    \label{fig:Fig7}
\end{figure}

Now we display the numerical results. In Table~\ref{tab:tab3} we show the results of $ZZ\gamma$ production at different $p^\gamma_{T,min}$. The ratio $\sigma^{gg}/\sigma^{q\Bar{q}}$ as a function of $p^\gamma_{T,min}$ is shown in Fig.~\ref{fig:Fig7}. From these results, we find that $\sigma^{gg}(ZZ\gamma)/\sigma^{q\Bar{q}}(ZZ\gamma)$ is about one order of magnitude smaller than $\sigma^{gg}(Z\gamma\gamma)/\sigma^{q\Bar{q}}(Z\gamma\gamma)$, which could also be explained by the gluon PDF. 

\begin{table}
    \renewcommand{\arraystretch}{1.5}
    \centering
    \scalebox{1.0}{
    \begin{tabular}{|c|c|c|}
    \hline
     $p^\gamma_{T,min}~$[GeV]  & $\sigma^{q\Bar{q}}(ZZ\gamma)~$[fb] & $\sigma^{gg}(ZZ\gamma)~$[$\times10^{-4}~$fb]\\
      \hline
     10  &  $46.9(2)^{+1.8\%}_{-2.3\%}$ & $25.10(5)^{+24.0\%}_{-18.4\%}$ \\
     20  &  $26.19(9)^{+1.2\%}_{-1.7\%}$ & $21.61(4)^{+24.3\%}_{-18.5\%}$ \\
     30  &  $17.31(7)^{+0.7\%}_{-1.2\%}$ & $17.78(7)^{+24.7\%}_{-18.8\%}$ \\
     40  &  $12.17(5)^{+0.4\%}_{-0.8\%}$ & $14.53(4)^{+25.2\%}_{-19.0\%}$ \\
     50  &  $9.05(4)^{+0.1\%}_{-0.5\%}$ & $11.80(5)^{+25.5\%}_{-19.2\%}$ \\
     60  &  $6.93(2)^{+0.0\%}_{-0.2\%}$ & $9.76(2)^{+25.9\%}_{-19.5\%}$ \\
     70  &  $5.42(2)^{+0.3\%}_{-0.6\%}$ & $8.15(1)^{+26.3\%}_{-19.7\%}$\\
     80  &  $4.35(2)^{+0.5\%}_{-0.8\%}$ & $6.84(1)^{+26.6\%}_{-19.8\%}$\\
     90  &  $3.53(1)^{+0.9\%}_{-1.1\%}$ & $5.82(1)^{+26.9\%}_{-20.0\%}$ \\
      \hline   
    \end{tabular}   }
    \caption{Total cross section for $ZZ\gamma$ production at the 13 TeV LHC with different $p^\gamma_{T,min}$, using the cuts of Eq.~(\ref{Eq1}) for photons. The results for $ZZ$ production are shown in Eq.~(\ref{eq2}).}
    \label{tab:tab3}
\end{table}

\subsection{\texorpdfstring{$W^+W^-\gamma$}{WWy} and \texorpdfstring{$W^+W^-$}{ww} productions}

\begin{figure}[htbp]%
    \centering
    \subfigure[]{
        \includegraphics[width=14cm]{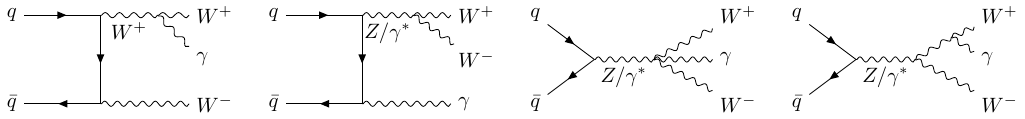}
        }\hfill
    \subfigure[]{
        \includegraphics[width=14cm]{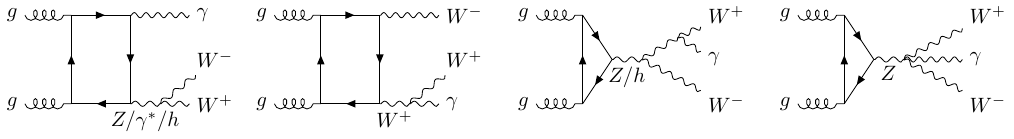}
        }\hfill
    \subfigure[]{
        \includegraphics[width=8cm]{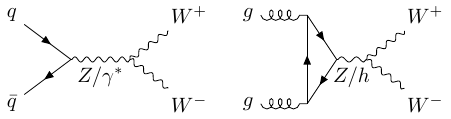}    
        }
    \caption{Additional Feynman diagrams in unitary gauge that contribute to (a) $q\Bar{q}\rightarrow W^+W^-\gamma$, (b) $gg\rightarrow W^+W^-\gamma$ and (c) $\{ q\Bar{q},~gg \} \rightarrow W^+W^-$ besides Fig.~\ref{fig:Fig1}.}
    \label{fig:Fig8}
\end{figure}

From the analysis in the preceding subsections, one may expect that the calculation for  $W^+W^-\gamma$ and $W^+W^-$ productions would be rather simple, which is not the case as shown in the following.  
We start with the cross section of  $W^+W^-$ production, which is given by 
\begin{equation}
    \label{eq4}
    \begin{aligned}
    \sigma^{q\Bar{q}}(W^+W^-)&=76.4(1)^{+3.6\%}_{-4.4\%}~\rm pb,\\
    \sigma^{gg}(W^+W^-)&=2.872(4)^{+21.5\%}_{-16.6\%}~\rm pb. 
    \end{aligned}
\end{equation}
Similar as $ZZ$ and $ZZ\gamma$ productions, there are new types of Feynman diagrams besides the ones plotted in Fig.~\ref{fig:Fig1}. 
These new Feynman diagrams contributing to $pp\rightarrow W^+W^-\gamma$ and $pp\rightarrow W^+W^-$ in the unitary gauge are shown in Fig.~\ref{fig:Fig8}. Besides the ordinary interactions which are already encountered in the previous examples, the triple and quartic gauge boson interactions also appear in these new diagrams. 
Now the complicated amplitude structures make it hard to get any conclusion about their total cross sections before numerical calculations. What we can only say is that if the contribution from Fig.~\ref{fig:Fig8} is neglected, the ratio $\sigma^{gg}/\sigma^{q\Bar{q}}$ for the triboson production should have the same order of magnitude as $ZZ\gamma$ production. Taking $\sigma^{q\Bar{q}}(W^+W^-)$ as an example, we find that there is a big cancellation between the $t$-channel and $s$-channel amplitudes through explicit calculations. Since similar cancellations also happen for the more complicated process $q\Bar{q}\rightarrow W^+W^-\gamma$ and the results of $W^+W^-\gamma$ will be shown explicitly later, here we skip the simple proof for $W^+W^-$. Before going to the numerical results, we want to emphasis that the subtracted amplitudes should not be gauge invariant and the corresponding cross sections which have no physical meanings are just used to understand the differences between the triboson processes discussed in this work. 

\begin{figure}
    \centering
    \includegraphics[width=11cm]{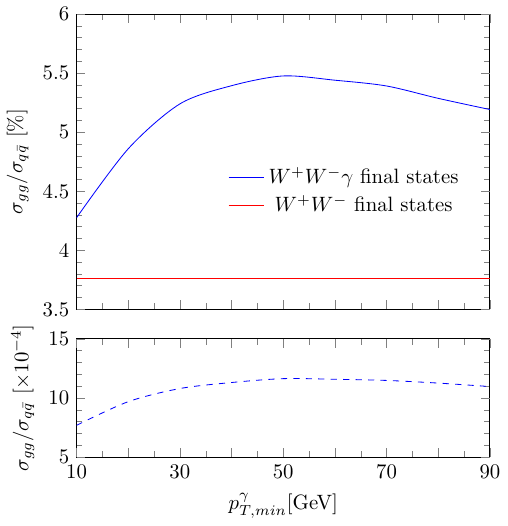}
    \caption{The ratio $\sigma_{gg}/\sigma_{q\Bar{q}}$ versus $p^\gamma_{T,min}$. In the upper panel we considered all diagrams shown in Fig.~\ref{fig:Fig1} and Fig.~\ref{fig:Fig8} for $W^+W^-\gamma$ and $W^+W^-$ productions. In the lower panel we only considered the diagrams in Fig.~\ref{fig:Fig1} for $W^+W^-\gamma$ production.}
    \label{fig:Fig9}
\end{figure}

The cross section for $W^+W^-\gamma$ production at different $p^\gamma_{T,min}$ is shown in Table~\ref{tab:tab4}. Here $\sigma_{F1}$ represents the contribution which only comes from the Feynman diagrams plotted in Fig.~\ref{fig:Fig1}. The ratio between gluon-gluon channel and quark-antiquark channel is shown in Fig.~\ref{fig:Fig9}. 
From these results we see that the ratio $\sigma^{gg}/\sigma^{q\Bar{q}}$ can reach 5\% due to the cancellation in $q\bar{q}\rightarrow W^+W^-\gamma$. Due to the large couplings between quarks and $W$-boson, the cross sections $\sigma_{F1}^{gg,q\bar{q}}$ are much larger than the corresponding cross sections of $ZZ\gamma$ production. On the other hand, for $\sigma_{F1}^{gg}/\sigma_{F1}^{q\bar{q}}$ we get the same order as for the process of $Z$-bosons as expected, since the ratio is insensitive to the interactions between quarks and gauge bosons.  

\begin{table}
    \renewcommand{\arraystretch}{1.5}
    \centering
    \scalebox{0.85}{
    \begin{tabular}{|c|c|c|c|c|}
    \hline
     $p^\gamma_{T,min}~$[GeV]  & $\sigma^{q\Bar{q}}(W^+W^-\gamma)~$[fb] & $\sigma^{q\Bar{q}}_{F1}(W^+W^-\gamma)~$[fb] & $\sigma^{gg}(W^+W^-\gamma)~$[fb] & $\sigma^{gg}_{F1}(W^+W^-\gamma)~$[fb] \\
      \hline
     10  & $300.8(9)^{+1.8\%}_{-2.4\%}$ & 6776 & $12.86(2)^{+23.7\%}_{-18.2\%}$ & 5.203 \\
     20  & $163.1(5)^{+1.1\%}_{-1.6\%}$ & 4698 & $7.925(8)^{+24.3\%}_{-18.5\%}$ & 4.548 \\ 
     30  & $105.3(4)^{+0.6\%}_{-1.0\%}$ & 3680 & $5.517(9)^{+24.7\%}_{-18.8\%}$ & 3.972 \\
     40  & $75.0(3)^{+0.1\%}_{-0.5\%}$ & 3046 & $4.045(5)^{+25.1\%}_{-19.0\%}$ & 3.443 \\
     50  & $56.3(2)^{+0.0\%}_{-0.2\%}$ & 2582 & $3.083(4)^{+25.4\%}_{-19.2\%}$ & 3.003 \\
     60  & $44.1(2)^{+0.3\%}_{-0.5\%}$ & 2262 & $2.399(3)^{+25.8\%}_{-19.4\%}$ & 2.619 \\
     70  & $35.3(1)^{+0.5\%}_{-0.8\%}$ & 2008 & $1.903(4)^{+25.9\%}_{-19.4\%}$ & 2.305 \\
     80  & $29.0(1)^{+0.8\%}_{-1.0\%}$ & 1801 & $1.533(2)^{+26.3\%}_{-19.7\%}$ & 2.027 \\
     90  & $24.07(9)^{+1.1\%}_{-1.3\%}$ & 1635 & $1.246(2)^{+26.6\%}_{-19.8\%}$ & 1.792 \\
      \hline   
    \end{tabular}   }
    \caption{Total cross sections for $W^+W^-\gamma$ production at different $p^\gamma_{T,min}$ in $pp$ collider, using the cuts of Eq.~(\ref{Eq1}) for photons. The subscript '$F1$' represents the unphysical cross sections only considering the Feynman diagrams in Fig.~\ref{fig:Fig1}(a). Thus the integration error and scale uncertainty are not provided for them. The results for the production of $W^+W^-$ are given in Eq.~(\ref{eq4}).}
    \label{tab:tab4}
\end{table}

From the experimental side, the measured fiducial cross section for $W^+W^-\gamma$ production with an integrated luminosity of 138  $\rm fb^{-1}$\cite{CMS:2023jpy} at the $13~\rm TeV$ LHC is in good agreement with the NLO QCD prediction. The relative experimental error is around 28\% and it surpasses the gluon-gluon channel contribution which is about 5\% of the LO value. 
At the High Luminosity LHC (HL-LHC) with $\sqrt{s}=14$ TeV and a luminosity of 3 $\rm ab^{-1}$, the experimental error could be reduced to a few percent. Meanwhile, as the increase of the center-of-mass energy, the contribution from the gluon-gluon channel will become more important. Thus the gluon-gluon channel contribution to $W^+W^-\gamma$ production should be considered in the future analysis. While for $Z\gamma\gamma$ and $ZZ\gamma$, the ratio  $\sigma_{gg}/\sigma_{q\Bar{q}}$ is much smaller and thus the gluon-gluon channel contribution could be safely neglected.

Some comments are in order before going to the conclusion. First of all, we  emphasis that it is not our aim to perform a precision study on the whole NNLO correction to the production of three gauge bosons at the LHC.  Actually, our focus is the pure gluon contribution which is gauge independent and might be sensitive to possible new physics beyond the SM. 
Thus our analysis could be useful for other BSM studies in the future from the viewpoint of concrete new physics models or effective theories. On the other hand, technically we do not just run the public code {\tt Madgraph} to get the physical results. 
To get the detailed information of the amplitudes, certain Feynman diagrams, which sometimes even are not physically gauge invariant by themselves, are picked out and then different operation is performed on the corresponding amplitude. 
For example, in Sec.~2.1 the new cross section $\sigma^*$ defined in $R(q)$ is obtained through replacing the numerical amplitudes with their absolute values. During our calculation, all such kinds of manipulations on the amplitudes are performed with the help of {\tt FeynArts}, {\tt FormCalc} and {\tt Collier}.

\section{Conclusion}
We calculated the gluon-gluon initiate state contribution to the productions of three gauge bosons at the 13 $\rm TeV$ LHC, which is formally part of NNLO effects compared  to the LO quark-antiquark channels. To understand the obtained results, the ratio between gluon-gluon channel contribution and the quark-antiquark channel contribution was presented and three different diboson production processes were chosen for comparative studies.  
We found that the ratio $\sigma_{gg}/\sigma_{q\Bar{q}}$ for $Z\gamma\gamma$ ($ZZ\gamma$) production is of the order of $10^{-3}$ ($10^{-4}$), much smaller than the corresponding ratio of diboson production due to the decrease of gluon PDF when more particles appear in the final states.  
These tiny ratios imply that gluon-gluon fusion contribution is phenomenological negligible for these two processes. However, for $W^+W^-\gamma$ production, the ratio $\sigma^{gg}/\sigma^{q\Bar{q}}$ can reach about 5\%,  at the same order of magnitude as the ratio for $W^+W^-$ because of the big cancellation between the amplitudes of quark-antiquark channels. Due to the large experimental uncertainty on the fiducial cross section, currently such gluon-gluon fusion effects can be safely neglected, while at the HL-LHC these effects may be accessible and should be considered. 
 
\section*{Acknowledgments}
This work was supported in part by IHEP under Grant No. Y9515570U1,
by the National Natural Science Foundation of China (NNSFC)
under grant Nos. 12375082, 12135013, 11821505, 12075300 and 12335005, by Peng-Huan-Wu Theoretical Physics Innovation
Center (12047503), by the CAS Center for Excellence in Particle Physics (CCEPP), and by the Key Research Program of the Chinese Academy of Sciences, Grant NO. XDPB15.

 \bibliographystyle{JHEP}
 \bibliography{biblio.bib}


\end{document}